\documentclass[aps,pre,twocolumn,superscriptaddress,showpacs]{revtex4-1}
\usepackage{amssymb,amsfonts,amsmath,amsthm}

\usepackage[dvips]{graphicx}
\usepackage{color}

\newcommand{\ZZ}{\mathbb{Z}}

\newcommand{\vzero}{\boldsymbol{0}}

\newcommand{\vk}{\boldsymbol{k}}

\newcommand{\veps}{\boldsymbol{\epsilon}}

\newcommand{\vphi}{\boldsymbol{\varphi}}

\begin{document}

\title{General approach for dealing with dynamical systems with spatiotemporal periodicities}

\author{Jes\'us Casado-Pascual}
\email{jcasado@us.es}
\affiliation{F\'{\i}sica Te\'orica,
Universidad de Sevilla, Apartado de Correos 1065, 41080 Sevilla, Spain}

\author{Jos\'e A.\ Cuesta}
\email{cuesta@math.uc3m.es}
\affiliation{Grupo Interdisciplinar de Sistemas Complejos (GISC), Departamento
de Matem\'aticas, Universidad Carlos III de Madrid, Avenida de la Universidad 30,
28911 Legan\'es, Spain}
\affiliation{Instituto de Biocomputaci\'on y F\'{\i}sica de Sistemas Complejos
(BIFI), Universidad de Zaragoza, 50009 Zaragoza, Spain}

\author{Niurka R.\ Quintero}
\email{niurka@us.es}
\affiliation{Instituto de Matem\'aticas de la Univesidad de Sevilla (IMUS)}
\affiliation{Departamento de F\'\i sica Aplicada I, E.P.S., Universidad de
Sevilla, Virgen de \'Africa 7, 41011, Sevilla, Spain}

\author{Renato Alvarez-Nodarse}
\email{ran@us.es}
\affiliation{Instituto de Matem\'aticas de la Univesidad de Sevilla (IMUS)}
\affiliation{Departamento de An\'alisis Matem\'atico, Universidad de Sevilla,
Apdo 1160, 41080, Sevilla, Spain}

\date{\today}

 
\begin{abstract} 
Dynamical systems often contain oscillatory forces or depend on periodic potentials. Time or space periodicity is reflected in the properties of these systems through a dependence on the parameters of their periodic terms. In this paper we provide a general theoretical framework for dealing with these kinds of systems, regardless of whether they are classical or quantum, stochastic or deterministic, dissipative or nondissipative, linear or nonlinear, etc. In particular, we are able to show that simple symmetry considerations determine, to a large extent, how their properties depend functionally on some of the parameters of the periodic terms. For the sake of illustration, we apply this formalism to find the functional dependence of the expectation value of the momentum of a Bose-Einstein condensate, described by the Gross-Pitaewskii equation, when it is exposed to a sawtooth potential whose amplitude is periodically modulated in time. We show that, by using this formalism, a small set of measurements is enough to obtain the functional form for a wide range of parameters. This can be very helpful when characterizing experimentally the response of systems for which performing measurements is costly or difficult.

\end{abstract}

%
%
\maketitle

It often happens that a system---physical or otherwise---can be described using
a model that includes one or several periodic functions, with the same or
different periodicities.  Whether these functions represent external
oscillatory forces, modulating amplitudes, or space-periodic potentials is
immaterial for our forthcoming discussion, as are the specific details of the
underlying dynamics, which can be either deterministic or stochastic, classical
or quantum, dissipative or nondissipative, linear or nonlinear, etc. For our purposes, the only relevant feature that all these
systems have in common is that their properties depend in a certain way on the
periodicities, amplitudes, and relative phases of these functions.

Examples of these types of systems abound in the literature. Because of their
ability to describe a wide variety of phenomena, the most significant ones are
probably the periodically driven systems
\cite{jung:1993,*grifoni:1998,*kohler:2005}.  To name but a few instances,
these driven systems have proven to be useful in the study of stochastic
resonance \cite{gammaitoni:1998,*casado:2003,*casado:2005}, vibrational
resonance
\cite{landa:2000,*casado:2004,*cubero1:2006,*casado:2007,*wickenbrock:2012},
classical and quantum stochastic synchronization
\cite{anishchenko:2002,*freund:2003,*lindner:2004,*casado1:2005,*goychuk:2006},
opinion formation processes
\cite{tessone:2005,*vaz_martins:2009,*tessone:2009}, coherent destruction of
tunneling \cite{grossmann:1991}, dynamical localization and delocalization
\cite{casati:1979,*moore:1994,*robinson:1995,*ringot:2000}, ratchet effect
\cite{astumian:2002,*reimann:2002a,*hanggi:2005,*hanggi:2009}, and atomic
quantum motors \cite{ponomarev:2009}. It is worth mentioning that some of the
above examples (e.g., the ratchet effect) include both time- and space-periodic
functions.

The purpose of this paper is to provide a general description of this
widespread situation, namely, a system whose dynamics depends on a set of
periodic functions, and to discuss the consequences of some symmetries that are
often encountered in this class of systems. Specifically, we show that the
functional dependence of the system's properties on some of the parameters contained in these functions may be determined to a large extent by
simple symmetry considerations.

To be more precise, let us suppose that the dynamics of the system under consideration depends
on $N$ periodic functions of a single variable $\zeta$. This variable may represent time, space, or even a generalized coordinate of some sort. As a matter of fact, our forthcoming discussion can be readily extended to the case in which there are more than one variable---say, space and time, or several spatial variables---, but for the sake of clarity we first consider only one. These periodic functions are assumed to be of the form
\begin{equation}
\label{eq1}
f_j(\zeta)=\alpha_j \cos[\Omega_j(\zeta-\zeta_0)]+\beta_j \sin[\Omega_j(\zeta-\zeta_0)]\,,
\end{equation}
with $j=1,\dots,N$, where $\Omega_j$ are the (temporal, spatial, or generalized) angular frequencies, and $\alpha_j$ and $\beta_j$ the partial amplitudes. This assumption is not as restrictive as it might appear, since any well-behaved periodic function can be approximated to any desired degree of accuracy by a finite sum of trigonometric functions. 
The parameter $\zeta_0$ has been introduced in Eq.~(\ref{eq1}) to simultaneously shift all the functions $f_j(\zeta)$ along the $\zeta$-axis. 

Let $\Upsilon$ represent a certain (physical) quantity of the system. We are interested in the functional dependence of $\Upsilon$ on some of the parameters appearing in the functions $f_j(\zeta)$. To study this dependence, we will make use of the following simple rule: any transformation of the parameters defining the functions $f_j(\zeta)$, which leaves these functions invariant, also leaves the value of the quantity $\Upsilon$ invariant. For this to be true, it is evidently assumed that all the other parameters in the problem remain unchanged during this transformation. 

In order to apply this rule, it is convenient to rewrite the periodic functions in Eq.~(\ref{eq1}) in the form $f_j(\zeta)=\epsilon_j \cos \left[\Omega_j(\zeta-\zeta_0)+\varphi_j\right]$, where we have introduced the $N$ amplitudes $\epsilon_j=(\alpha_j^2+\beta_j^2)^{1/2}$ and the $N$ phases $\varphi_j$, satisfying the equations $\cos \varphi_j=\alpha_j/\epsilon_j$ and  $\sin \varphi_j=-\beta_j/\epsilon_j$. Let us define the vectors $\boldsymbol{\Omega}=(\Omega_1,\dots,\Omega_N)$, $\boldsymbol{\varphi}=(\varphi_1,\dots,\varphi_N)$, and $\boldsymbol{\epsilon}=(\epsilon_1,\dots,\epsilon_N)$. Then, it is clear that the set of periodic functions is invariant under these two transformations 
\begin{eqnarray}
\label{trans1}
\mathcal{T}_1&:& \{\zeta_0,\boldsymbol{\Omega},\vphi,\veps\}\longmapsto\{\zeta_0,\boldsymbol{\Omega},\vphi+\pi \boldsymbol{u}^{(j)},\veps^{(j)}\}\,,\\
\label{trans2}
\mathcal{T}_2&:& \{\zeta_0,\boldsymbol{\Omega},\vphi,\veps\}\longmapsto\{0,\boldsymbol{\Omega},\vphi-\zeta_0 \boldsymbol{\Omega},\veps\}\,,
\end{eqnarray}
where $\boldsymbol{u}^{(j)}$ denotes the $j$th row of the $N\times N$ identity matrix and the vector $\veps^{(j)}$ is obtained from the vector $\veps$ by replacing its $j$th component by $-\epsilon_j$. Consequently, from the above-mentioned rule, 
\begin{equation}
\label{eq2}
\Upsilon(\zeta_0,\vphi,\veps)=\Upsilon(\zeta_0,\vphi+\pi\boldsymbol{u}^{(j)},\veps^{(j)})
=\Upsilon(0,\vphi-\boldsymbol{\Omega}\zeta_0,\veps)\,,
\end{equation}
where we have only explicitly written the dependence of $\Upsilon$ on the parameters $\zeta_0$, $\vphi$, and $\veps$, for $\boldsymbol{\Omega}$ is assumed to be fixed throughout this study. The analysis of the dependence on $\boldsymbol{\Omega}$ requires the use of alternative techniques \cite{casado:2013,*cubero:2014}.  

By applying the first equality in the above equation twice, we see that $\Upsilon(\zeta_0,\vphi,\veps)$ is periodic with respect to all the components of the vector $\vphi$ with period $2\pi$. Therefore, taking into account the second equality in Eq.~(\ref{eq2}), it can be expanded in Fourier series as 
\begin{equation}
\label{Fourier1}
\Upsilon(\zeta_0,\vphi,\veps)=\sum_{\vk\in\ZZ^N}\upsilon_{\vk}(\veps)
e^{i (\vphi-\boldsymbol{\Omega} \zeta_0)\cdot \vk}\,,
\end{equation}
where $\boldsymbol{\varphi}\cdot \vk=\sum_{j=1}^N \varphi_j k_j$ and
\begin{equation}
\label{eq3}
\upsilon_{\vk}(\veps)=\int_{-\pi}^{\pi}\dots\int_{-\pi}^{\pi} \frac{d^N\vphi}{(2\pi)^N}\,\Upsilon(0,\vphi,\veps)e^{-i\vphi\cdot \vk}\,.
\end{equation}

Without loss of generality, we can assume that the quantity $\Upsilon$ is real. Otherwise, one would consider its real and imaginary parts separately. Then, taking into account that the imaginary part
of Eq.~(\ref{Fourier1}) is zero and introducing the functions $\gamma_{\vk}(\veps)=\upsilon_{\vk}(\veps)\prod_{j=1}^N\epsilon_j^{-|k_j|}$, one obtains
\begin{multline}
\label{Fourier2}
\Upsilon(\zeta_0,\vphi,\veps)=\sum_{\vk\in\ZZ^N}|\gamma_{\vk}(\veps)|\Bigg(\prod_{j=1}^N\epsilon_j^{|k_j|}\Bigg)\\
\times  \cos\left[(\vphi-\boldsymbol{\Omega} \zeta_0)\cdot \vk+\chi_{\vk}(\veps)\right]\,,
\end{multline}
where $|\gamma_{\vk}(\veps)|$ and $\chi_{\vk}(\veps)$ are, respectively, the modulus and phase of the complex number $\gamma_{\vk}(\veps)$. Note that, according to Eq.~(\ref{eq3}) and the first equality in Eq.~(\ref{eq2}), the functions $\gamma_{\vk}(\veps)$ and, consequently, $|\gamma_{\vk}(\veps)|$ and $\chi_{\vk}(\veps)$, are even in each of the arguments $\epsilon_j$. In addition, since $\Upsilon$ is real, $\left|\gamma_{\boldsymbol{k}}(\boldsymbol{\epsilon})\right|=\left|\gamma_{-\boldsymbol{k}}(\boldsymbol{\epsilon})\right|$ and $e^{i\chi_{\boldsymbol{k}}(\boldsymbol{\epsilon})}=e^{-i\chi_{-\boldsymbol{k}}(\boldsymbol{\epsilon})}$.

An important result follows from assuming that the quantity $\Upsilon$ is invariant under arbitrary shifts of all the periodic functions $f_j(\zeta)$ along the $\zeta$-axis. In this case, $\Upsilon$ is independent of $\zeta_0$ and, accordingly, all the
coefficients $\gamma_{\vk}(\veps)$ such that $\vk\cdot\boldsymbol{\Omega}\ne 0$ must be zero. Thus, dropping the dependence of $\Upsilon(\zeta_0,\vphi,\veps)$ on $\zeta_0$, we obtain
\begin{equation}
\label{Fourier3}
\Upsilon(\vphi,\veps)=\sum_{\vk\in\mathcal{S}_{\boldsymbol{\Omega}}}|\gamma_{\vk}(\veps)|
\cos\left[\vphi\cdot \vk+\chi_{\vk}(\veps)\right]\prod_{j=1}^N\epsilon_j^{|k_j|}\,,
\end{equation}
where $\mathcal{S}_{\boldsymbol{\Omega}}$ is the set of all ordered $N$-tuples of integers orthogonal to $\boldsymbol{\Omega}$, i.e.,  $\mathcal{S}_{\boldsymbol{\Omega}}:=\{\vk\in\ZZ^N\,:\,\boldsymbol{\Omega}\cdot \vk=0\}$.
It is worth noting that, independently of whether or not $\Upsilon$ satisfies the above-mentioned shift-invariance property, Eq.~(\ref{Fourier3}) is always valid for the average value $\bar{\Upsilon}(\vphi,\veps)=\lim_{\Delta\zeta_0\to\infty}(\Delta\zeta_0)^{-1}\int_0^{\Delta\zeta_0}d\zeta_0\,\Upsilon(\zeta_0,\vphi,\veps)$.

Let us now consider the case in which the $N$ angular frequencies $\Omega_1,\dots,\Omega_N$ are incommensurable, i.e., it is not possible to express one of them as a linear combination of the others with rational coefficients. Then, the set $\mathcal{S}_{\boldsymbol{\Omega}}$ consists of the single element $\vk=\vzero$ and, according to Eq.~(\ref{Fourier3}), $\Upsilon(\vphi,\veps)=\gamma_{\vzero}(\veps)$. Consequently, the quantity $\Upsilon$ is an even function in each of the components of the vector $\veps$ and is independent of the phases $\vphi$. If there were additional symmetry considerations leading to the conclusion that $\Upsilon$ is an odd function in any of the components of $\veps$, then necessarily $\Upsilon=0$ (see, e.g., Refs.\cite{neumann:2002,cubero:2012} for the case of two incommensurable frequencies). 

In contrast, if the $N$ angular frequencies are pairwise commensurable, then there exists a frequency $\bar{\Omega}$ such that $\boldsymbol{\Omega}=\bar{\Omega} \boldsymbol{n}$, where $\boldsymbol{n}=(n_1,\dots,n_N)$, with $n_j$ being positive integers.
Thus, the condition $\boldsymbol{\Omega}\cdot \vk=0$ is equivalent to the Diophantine equation $\boldsymbol{n}\cdot\vk=0$. The general solution of this last equation can be expressed as an integer linear combinations of a set of $N-1$ generating vectors, $\vk^{(1)},\dots,\vk^{(N-1)}$, each of which satisfies $\boldsymbol{n}\cdot\vk^{(j)}=0$
\cite{morito:1980}. Consequently, Eq.~(\ref{Fourier3}) now reads
\begin{multline}
\label{Fourier4}
\Upsilon(\vphi,\veps)=\sum_{\mathbf{q}\in \ZZ^{N-1}} |\gamma_{\vk(\mathbf{q})}(\veps)|\left(\prod_{j=1}^N\epsilon_j^{|k_j(\mathbf{q})|}\right)\\
\times \cos\left[\vphi\cdot\vk(\mathbf{q})+\chi_{\vk(\mathbf{q})}(\veps)\right]\,,
\end{multline}
where $\vk(\mathbf{q})=\sum_{l=1}^{N-1} q_l \vk^{(l)}$. Hence, $\Upsilon$ depends on $\vphi$ only through the
collective phases $\vphi\cdot\vk^{(1)},\dots,\vphi\cdot\vk^{(N-1)}$. Equation~(\ref{Fourier4}) can be considered as a generalization of the results reported in Ref.\cite{cuesta:2013} in the context of rocking ratchets.

We now proceed to study the perturbative behavior of the quantity $\Upsilon$ for sufficiently small values of the amplitudes $\epsilon_j$, expressed in suitable dimensionless units. To this end, we will assume that, for $\zeta_0=0$, the quantity $\Upsilon$, expressed as a function of the partial amplitudes $\boldsymbol{\alpha}$ and $\boldsymbol{\beta}$, has a Taylor power expansion of the form  $\sum_{\boldsymbol{l},\boldsymbol{r}\in \mathbb{N}_0^N} a_{\boldsymbol{l},\boldsymbol{r}} \prod_{j=1}^N\alpha_j^{l_j}\beta_j^{r_j}$, with $\mathbb{N}_0$ being the set of all nonnegative integers and $a_{\boldsymbol{l},\boldsymbol{r}}$ the coefficients of the Taylor series. Then, using that $\alpha_j=\epsilon_j \cos\varphi_j$ and $\beta_j=-\epsilon_j \sin\varphi_j$, as well as Eq.~(\ref{eq3}) and the definition of the functions $\gamma_{\vk}(\veps)$, we find
\begin{equation}
\gamma_{\vk}(\veps)=\sum_{\boldsymbol{l},\boldsymbol{r}\in \mathbb{N}_0^N} a_{\boldsymbol{l},\boldsymbol{r}}\prod_{j=1}^N\epsilon_j^{l_j+r_j-|k_j|} I_{l_j,r_j,k_j}\,,
\end{equation}
where
\begin{equation}
I_{l,r,k}=\int_{-\pi}^{\pi}\frac{d \varphi}{2\pi}\,\cos^l\varphi\sin^r\varphi\,e^{-i  \varphi k}\,.
\end{equation}
The above integral vanishes unless $l+r-|k|$ is a nonnegative even integer. Consequently,
\begin{equation}
\label{perturbative}
\gamma_{\vk}(\veps)=\sum_{\boldsymbol{p}\in \mathbb{N}_0^N} b_{\boldsymbol{k},\boldsymbol{p}} \prod_{j=1}^N \epsilon_j^{2 p_j}\,,
\end{equation}
with 
\begin{equation}
b_{\boldsymbol{k},\boldsymbol{p}}=\sum_{\boldsymbol{l},\boldsymbol{r}\in \mathbb{N}_0^N} a_{\boldsymbol{l},\boldsymbol{r}}\prod_{j=1}^N I_{l_j,r_j,k_j} \delta_{2p_j,l_j+r_j-|k_j|}\,.
\end{equation}
The series expansion in Eq.~(\ref{perturbative}), together with Eqs.~(\ref{Fourier2}), (\ref{Fourier3}), or (\ref{Fourier4}), depending on the specific case, allows determining the functional dependence of the quantity $\Upsilon$ on the parameters $\zeta_0$, $\veps$, and $\vphi$. In practice, for sufficiently small amplitudes, these expansions can be truncated to include only a few terms. In this case, the determination of $\Upsilon(\zeta_0,\vphi,\veps)$ is reduced to the calculation of a few model-dependent coefficients $b_{\boldsymbol{k},\boldsymbol{p}}$.

In order to understand how to put these ideas into practice, let us consider a one-dimensional Bose-Einstein condensate described by the nonlinear Gross-Pitaevskii equation~\cite{dalfovo:1999,*pitaevskii:2003}
\begin{multline}
\label{GP}
i\hbar \frac{\partial{\Psi(x,t)}}{{\partial t}}=-\frac{\hbar^2}{2m}\frac{\partial^2\Psi(x,t)}{\partial x^2}+U(x,t)\Psi(x,t)\\
+g |\Psi(x,t)|^2\Psi(x,t)\,,
\end{multline}
where $\Psi(x,t)$ is the condensate wave function, normalized to 1, $m$ the mass of the bosons, $g$ the scaled strength of the nonlinear interaction, and $U(x,t)$ a potential of the form $U(x,t)=U_0 V(x)\left[1+\cos(2\pi t/T)\right]$, with $U_0$ being a constant with the dimensions of energy and $T$ the temporal period. The spatial part $V(x)$ is described by a biharmonic function of the form $V(x)=\cos(2 \pi x/\lambda)+\eta \cos(4\pi  x/\lambda+\phi)$~\cite{poletti:2009a}, where $\eta$ is the amplitude of the second harmonic relative to that of the fundamental, $\lambda$ the spatial period, and $\phi$ the relative phase between the two harmonics. In order to numerically solve Eq.~(\ref{GP}), we consider an initial condition of the form $\Psi(x,0)=\sqrt{2/\lambda}\cos(2\pi x/\lambda)$, and impose periodic boundary conditions~\cite{poletti:2009a}, i.e., $\Psi(x+\lambda,t)=\Psi(x,t)$ $\forall x \in \mathbb{R}$. In this paper, we focus our attention on the expectation value of the momentum evaluated at time $t=T$, which is given by the expression
\begin{equation}
\label{pdefinition}
P_{T}=\int_{-\lambda/2}^{\lambda/2} dx\,\Psi^{*}(x,T)\left(-i \hbar \frac{\partial }{\partial x}\right)\Psi(x,T)\,.
\end{equation}
Specifically, we are interested in the functional dependence of $P_{T}$ on the parameters $\phi$ and $\eta$, i.e., in the function $P_{T}(\phi,\eta)$. 

To obtain an expression for $P_{T}(\phi,\eta)$, we first apply the previously developed formalism to the periodic functions $f_1(x)= \cos(2 \pi x/\lambda)$ and $f_2(x)=\eta\cos(4 \pi x/\lambda+\phi)$. The correspondence with our previous notation is $\zeta=x$, $\zeta_0=0$, $\boldsymbol{\Omega}=(2\pi/\lambda,4\pi/\lambda)$, $\vphi=(0,\phi)$, and $\veps=(1,\eta)$. Consequently, using Eq.~(\ref{Fourier2}) together with  Eq.~(\ref{perturbative}) leads to
\begin{equation} 
\label{result_1}
P_T(\phi,\eta)=\sum_{k,p=0}^{\infty}\eta^{k+2p}
\left[\mu_{k,p} \sin(k\phi) 
+\nu_{k,p}  \cos(k \phi)\right]\,,
\end{equation} 
where $\mu_{k,p}$ and $\nu_{k,p}$ are coefficients that do not depend on $\phi$ and $\eta$. Since Eq.~(\ref{GP}) as well as the considered initial condition are invariant under the transformation  $\mathcal{T}_3: \{x,\phi\} \longmapsto \{-x,-\phi\}$, it is easy to see that $P_T(-\phi,\eta)=-P_T(\phi,\eta)$. Hence, all the coefficients $\nu_{k,p}$ in Eq.~(\ref{result_1}) must vanish. 

Let us now assume that $\eta$ is sufficiently small so that we can neglect the terms of order $\mathcal{O}(\eta^4)$ in Eq.~(\ref{result_1}), and approximate $P_T(\phi,\eta)$ by
\begin{equation}
\label{result_2}
P_T(\phi,\eta) \approx \eta\left(\mu_{1,0}+\eta^2\mu_{1,1}\right)\sin \phi+\eta^2\mu_{2,0}\sin(2\phi)\,.
\end{equation}
The determination of the function $P_T(\phi,\eta)$ is thus reduced to evaluating the three coefficients $\mu_{1,0}$, $\mu_{1,1}$, and $\mu_{2,0}$. These coefficients can be easily calculated if we know, e.g., $P_T(\pi/2,\eta_1)$, $P_T(\pi/2,\eta_2)$, and $P_T(\pi/4,\eta_1)$, with $\eta_1$ and $\eta_2$ being two different values of $\eta$ within the validity range of Eq.~(\ref{result_2}). In that case,
\begin{equation}
\label{mu10}
\mu_{1,0}=\frac{\eta_2^3 P_T(\pi/2,\eta_1)-\eta_1^3 P_T(\pi/2,\eta_2)}{\eta_1\eta_2(\eta_2^2-\eta_1^2)}\,,
\end{equation}
\begin{equation}
\label{mu11}
\mu_{1,1}=\frac{\eta_2 P_T(\pi/2,\eta_1)-\eta_1 P_T(\pi/2,\eta_2)}{\eta_1\eta_2(\eta_1^2-\eta_2^2)}\,,
\end{equation}
and
\begin{equation}
\label{mu20}
\mu_{2,0}=\frac{2P_T(\pi/4,\eta_1)-\sqrt{2}\,\eta_1\left(\mu_{1,0}+\eta_1^2\mu_{1,1}\right)}{2\eta_1^2}\,.
\end{equation}
To sum up, it is enough to know $P_T$ for three different values of $(\phi,\eta)$ to determine its value for a wide range of parameters $\phi$ and $\eta$.

\begin{figure}
\begin{center}
\includegraphics[scale=0.75]{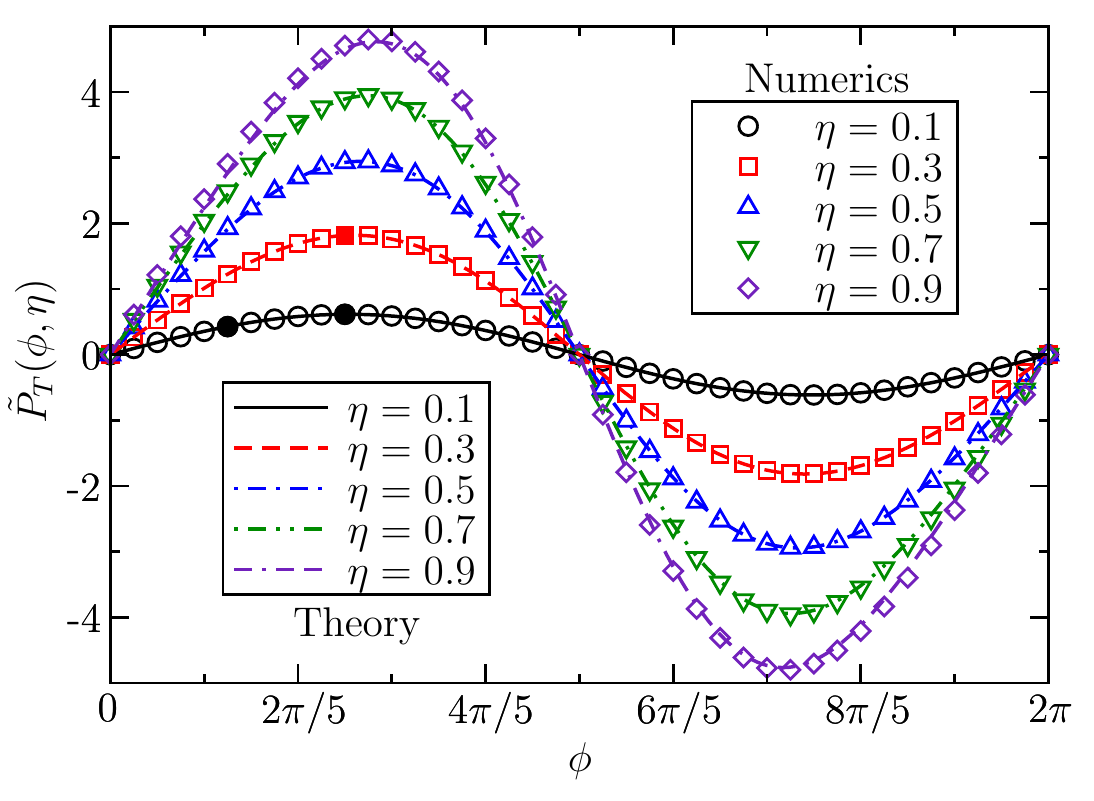}
\end{center}
\caption{(Color online) Dependence of the dimensionless momentum  $\tilde{P}_T(\phi,\eta)= {\hbar}^{-1}\lambda P_T(\phi,\eta)$ on the relative phase $\phi$ for $\eta=0.1$, $0.3$, $0.5$, $0.7$, and $0.9$. The rest of the parameter values, expressed in dimensionless form, are $\hbar^{-1} \lambda^2 T^{-1}m=1$, $\hbar^{-1} T U_0=1$, and $\hbar^{-1}\lambda^{-1}T g=1$. The results obtained from the numerical solution of the Gross-Pitaevskii equation in Eq.~(\ref{GP}) are represented by five different types of symbols, as shown in the upper legend box. The three solid symbols indicate the values used to evaluate the coefficients $\mu_{1,0}$, $\mu_{1,1}$, and $\mu_{2,0}$ from Eqs.~(\ref{mu10}), (\ref{mu11}), and (\ref{mu20}). The results calculated by using Eq.~(\ref{result_2}) with the obtained coefficients are depicted with five different types of lines, as indicated in the lower legend box.}
\label{fig1}
\end{figure}

In order to illustrate this result, we have used a spectral method~\cite{num_rec} to numerically solve Eq.~(\ref{GP}) for $(\phi,\eta)=(\pi/2,0.1)$, $(\pi/2,0.3)$, and $(\pi/4,0.1)$. The rest of the parameter values, conveniently expressed in dimensionless form, are $\hbar^{-1} \lambda^2 T^{-1}m=1$, $\hbar^{-1} T U_0=1$, and $\hbar^{-1}\lambda^{-1}T g=1$. The results obtained for the dimensionless momentum $\tilde{P}_T(\phi,\eta)= \hbar^{-1}\lambda P_T(\phi,\eta)$ are $\tilde{P}_T(\pi/2,0.1)\approx 0.6145$, $\tilde{P}_T(\pi/2,0.3)\approx 1.8143$, and $\tilde{P}_T(\pi/4,0.1)\approx 0.4281$. These three values are shown by solid symbols in Fig.~\ref{fig1}. By substituting these three values of $\tilde{P}_T(\phi,\eta)$ into Eqs.~(\ref{mu10}), (\ref{mu11}), and (\ref{mu20}), one can explicitly calculate the coefficients that appear in Eq.~(\ref{result_2}), yielding $\hbar^{-1}\lambda \mu_{1,0}\approx 6.1571$, $\hbar^{-1}\lambda \mu_{1,1}\approx -1.2083$, and $\hbar^{-1}\lambda \mu_{2,0}\approx -0.6417$. The results for $\tilde{P}_T(\phi,\eta)$ obtained by using Eq.~(\ref{result_2}) and these three values of the coefficients are depicted with five different types of lines in Fig.~\ref{fig1}. To check the accuracy of these predictions, we also show in Fig.~\ref{fig1}, with five different types of symbols, the results obtained from the numerical solution of the Gross-Pitaevskii equation in Eq.~(\ref{GP}). As can be seen, the agreement between our predictions and the numerical results is excellent.

In conclusion, we have developed a general theoretical framework for describing dynamical systems that contain periodic terms. The formalism can be equally applied whether the system is classical or quantum, stochastic or deterministic, 
dissipative or nondissipative, linear or nonlinear, and more importantly, regardless of whether the periodic terms are time oscillations (in external forces or modulating amplitudes) or periodic spatial potentials (as in the example), or both. We have shown that the functional dependence of the system's properties on some of the parameters of the periodic terms can be determined, to a large extent, by simple symmetry considerations. In particular, within the appropriate range of parameters, this functional dependence can be obtained, except for a few unknown constant coefficients. This last result can be very helpful when characterizing experimentally the response of systems for which performing measurements is costly or difficult.

\begin{acknowledgments} We acknowledge financial support through Grants No.
MTM2012-36732-C03-03 (R.A.N.), No. FIS2011-24540 (N.R.Q.), and PRODIEVO (J.A.C.),
from the Ministerio de Econom\'{\i}a y Competitividad (Spain), Grants No. FQM262
(R.A.N.), No. FQM207 (N.R.Q.), and Nos. FQM-7276, P09-FQM-4643 (N.R.Q., R.A.N.), from
the Ministerio de Ciencia e Innovaci\'on of Spain, Grant No. FIS2008-02873
(J.C.-P.), from Junta de Andaluc\'{\i}a (Spain), and from Alexander von Humboldt-Stiftung, Germany, through Research Fellowship for Experienced Researchers SPA, Grant No. 1146358
STP (N.R.Q.).  \end{acknowledgments}

\bibstyle{revtex} \bibliography{biblio}

\begin{thebibliography}{39}%
\makeatletter
\providecommand \@ifxundefined [1]{%
 \@ifx{#1\undefined}
}%
\providecommand \@ifnum [1]{%
 \ifnum #1\expandafter \@firstoftwo
 \else \expandafter \@secondoftwo
 \fi
}%
\providecommand \@ifx [1]{%
 \ifx #1\expandafter \@firstoftwo
 \else \expandafter \@secondoftwo
 \fi
}%
\providecommand \natexlab [1]{#1}%
\providecommand \enquote  [1]{``#1''}%
\providecommand \bibnamefont  [1]{#1}%
\providecommand \bibfnamefont [1]{#1}%
\providecommand \citenamefont [1]{#1}%
\providecommand \href@noop [0]{\@secondoftwo}%
\providecommand \href [0]{\begingroup \@sanitize@url \@href}%
\providecommand \@href[1]{\@@startlink{#1}\@@href}%
\providecommand \@@href[1]{\endgroup#1\@@endlink}%
\providecommand \@sanitize@url [0]{\catcode `\\12\catcode `\$12\catcode
  `\&12\catcode `\#12\catcode `\^12\catcode `\_12\catcode `\%12\relax}%
\providecommand \@@startlink[1]{}%
\providecommand \@@endlink[0]{}%
\providecommand \url  [0]{\begingroup\@sanitize@url \@url }%
\providecommand \@url [1]{\endgroup\@href {#1}{\urlprefix }}%
\providecommand \urlprefix  [0]{URL }%
\providecommand \Eprint [0]{\href }%
\providecommand \doibase [0]{http://dx.doi.org/}%
\providecommand \selectlanguage [0]{\@gobble}%
\providecommand \bibinfo  [0]{\@secondoftwo}%
\providecommand \bibfield  [0]{\@secondoftwo}%
\providecommand \translation [1]{[#1]}%
\providecommand \BibitemOpen [0]{}%
\providecommand \bibitemStop [0]{}%
\providecommand \bibitemNoStop [0]{.\EOS\space}%
\providecommand \EOS [0]{\spacefactor3000\relax}%
\providecommand \BibitemShut  [1]{\csname bibitem#1\endcsname}%
\let\auto@bib@innerbib\@empty
\bibitem [{\citenamefont {Jung}(1993)}]{jung:1993}%
  \BibitemOpen
  \bibfield  {author} {\bibinfo {author} {\bibfnamefont {P.}~\bibnamefont
  {Jung}},\ }\href@noop {} {\bibfield  {journal} {\bibinfo  {journal} {Phys.
  Rep.}\ }\textbf {\bibinfo {volume} {234}},\ \bibinfo {pages} {175} (\bibinfo
  {year} {1993})}\BibitemShut {NoStop}%
\bibitem [{\citenamefont {Grifoni}\ and\ \citenamefont
  {H{\"a}nggi}(1998)}]{grifoni:1998}%
  \BibitemOpen
  \bibfield  {author} {\bibinfo {author} {\bibfnamefont {M.}~\bibnamefont
  {Grifoni}}\ and\ \bibinfo {author} {\bibfnamefont {P.}~\bibnamefont
  {H{\"a}nggi}},\ }\href@noop {} {\bibfield  {journal} {\bibinfo  {journal}
  {Phys. Rep.}\ }\textbf {\bibinfo {volume} {304}},\ \bibinfo {pages} {229}
  (\bibinfo {year} {1998})}\BibitemShut {NoStop}%
\bibitem [{\citenamefont {Kohler}\ \emph {et~al.}(2005)\citenamefont {Kohler},
  \citenamefont {Lehmann},\ and\ \citenamefont {H{\"a}nggi}}]{kohler:2005}%
  \BibitemOpen
  \bibfield  {author} {\bibinfo {author} {\bibfnamefont {S.}~\bibnamefont
  {Kohler}}, \bibinfo {author} {\bibfnamefont {J.}~\bibnamefont {Lehmann}}, \
  and\ \bibinfo {author} {\bibfnamefont {P.}~\bibnamefont {H{\"a}nggi}},\
  }\href@noop {} {\bibfield  {journal} {\bibinfo  {journal} {Phys. Rep.}\
  }\textbf {\bibinfo {volume} {406}},\ \bibinfo {pages} {379} (\bibinfo {year}
  {2005})}\BibitemShut {NoStop}%
\bibitem [{\citenamefont {Gammaitoni}\ \emph {et~al.}(1998)\citenamefont
  {Gammaitoni}, \citenamefont {H{\"a}nggi}, \citenamefont {Jung},\ and\
  \citenamefont {Marchesoni}}]{gammaitoni:1998}%
  \BibitemOpen
  \bibfield  {author} {\bibinfo {author} {\bibfnamefont {L.}~\bibnamefont
  {Gammaitoni}}, \bibinfo {author} {\bibfnamefont {P.}~\bibnamefont
  {H{\"a}nggi}}, \bibinfo {author} {\bibfnamefont {P.}~\bibnamefont {Jung}}, \
  and\ \bibinfo {author} {\bibfnamefont {F.}~\bibnamefont {Marchesoni}},\
  }\href@noop {} {\bibfield  {journal} {\bibinfo  {journal} {Rev. Mod. Phys.}\
  }\textbf {\bibinfo {volume} {70}},\ \bibinfo {pages} {223} (\bibinfo {year}
  {1998})}\BibitemShut {NoStop}%
\bibitem [{\citenamefont {Casado-Pascual}\ \emph {et~al.}(2003)\citenamefont
  {Casado-Pascual}, \citenamefont {G{\'o}mez-Ord{\'o}{\~n}ez}, \citenamefont
  {Morillo},\ and\ \citenamefont {H{\"a}nggi}}]{casado:2003}%
  \BibitemOpen
  \bibfield  {author} {\bibinfo {author} {\bibfnamefont {J.}~\bibnamefont
  {Casado-Pascual}}, \bibinfo {author} {\bibfnamefont {J.}~\bibnamefont
  {G{\'o}mez-Ord{\'o}{\~n}ez}}, \bibinfo {author} {\bibfnamefont
  {M.}~\bibnamefont {Morillo}}, \ and\ \bibinfo {author} {\bibfnamefont
  {P.}~\bibnamefont {H{\"a}nggi}},\ }\href@noop {} {\bibfield  {journal}
  {\bibinfo  {journal} {Phys. Rev. Lett.}\ }\textbf {\bibinfo {volume} {91}},\
  \bibinfo {pages} {210601} (\bibinfo {year} {2003})}\BibitemShut {NoStop}%
\bibitem [{\citenamefont {Casado-Pascual}\ \emph
  {et~al.}(2005{\natexlab{a}})\citenamefont {Casado-Pascual}, \citenamefont
  {G{\'o}mez-Ord{\'o}{\~n}ez},\ and\ \citenamefont {Morillo}}]{casado:2005}%
  \BibitemOpen
  \bibfield  {author} {\bibinfo {author} {\bibfnamefont {J.}~\bibnamefont
  {Casado-Pascual}}, \bibinfo {author} {\bibfnamefont {J.}~\bibnamefont
  {G{\'o}mez-Ord{\'o}{\~n}ez}}, \ and\ \bibinfo {author} {\bibfnamefont
  {M.}~\bibnamefont {Morillo}},\ }\href@noop {} {\bibfield  {journal} {\bibinfo
   {journal} {Chaos}\ }\textbf {\bibinfo {volume} {15}},\ \bibinfo {pages}
  {026115} (\bibinfo {year} {2005}{\natexlab{a}})}\BibitemShut {NoStop}%
\bibitem [{\citenamefont {Landa}\ and\ \citenamefont
  {McClintock}(2000)}]{landa:2000}%
  \BibitemOpen
  \bibfield  {author} {\bibinfo {author} {\bibfnamefont {P.}~\bibnamefont
  {Landa}}\ and\ \bibinfo {author} {\bibfnamefont {P.~V.~E.}\ \bibnamefont
  {McClintock}},\ }\href@noop {} {\bibfield  {journal} {\bibinfo  {journal} {J.
  Phys. A}\ }\textbf {\bibinfo {volume} {33}},\ \bibinfo {pages} {L433}
  (\bibinfo {year} {2000})}\BibitemShut {NoStop}%
\bibitem [{\citenamefont {Casado-Pascual}\ and\ \citenamefont
  {Baltan\'as}(2004)}]{casado:2004}%
  \BibitemOpen
  \bibfield  {author} {\bibinfo {author} {\bibfnamefont {J.}~\bibnamefont
  {Casado-Pascual}}\ and\ \bibinfo {author} {\bibfnamefont {J.~P.}\
  \bibnamefont {Baltan\'as}},\ }\href@noop {} {\bibfield  {journal} {\bibinfo
  {journal} {Phys. Rev. E}\ }\textbf {\bibinfo {volume} {69}},\ \bibinfo
  {pages} {046108} (\bibinfo {year} {2004})}\BibitemShut {NoStop}%
\bibitem [{\citenamefont {Cubero}\ \emph {et~al.}(2006)\citenamefont {Cubero},
  \citenamefont {Baltan\'as},\ and\ \citenamefont
  {Casado-Pascual}}]{cubero1:2006}%
  \BibitemOpen
  \bibfield  {author} {\bibinfo {author} {\bibfnamefont {D.}~\bibnamefont
  {Cubero}}, \bibinfo {author} {\bibfnamefont {J.~P.}\ \bibnamefont
  {Baltan\'as}}, \ and\ \bibinfo {author} {\bibfnamefont {J.}~\bibnamefont
  {Casado-Pascual}},\ }\href@noop {} {\bibfield  {journal} {\bibinfo  {journal}
  {Phys. Rev. E}\ }\textbf {\bibinfo {volume} {73}},\ \bibinfo {pages} {061102}
  (\bibinfo {year} {2006})}\BibitemShut {NoStop}%
\bibitem [{\citenamefont {Casado-Pascual}\ \emph {et~al.}(2007)\citenamefont
  {Casado-Pascual}, \citenamefont {Cubero},\ and\ \citenamefont
  {Baltan\'as}}]{casado:2007}%
  \BibitemOpen
  \bibfield  {author} {\bibinfo {author} {\bibfnamefont {J.}~\bibnamefont
  {Casado-Pascual}}, \bibinfo {author} {\bibfnamefont {D.}~\bibnamefont
  {Cubero}}, \ and\ \bibinfo {author} {\bibfnamefont {J.~P.}\ \bibnamefont
  {Baltan\'as}},\ }\href@noop {} {\bibfield  {journal} {\bibinfo  {journal}
  {Europhys. Lett.}\ }\textbf {\bibinfo {volume} {77}},\ \bibinfo {pages}
  {50004} (\bibinfo {year} {2007})}\BibitemShut {NoStop}%
\bibitem [{\citenamefont {Wickenbrock}\ \emph {et~al.}(2012)\citenamefont
  {Wickenbrock}, \citenamefont {Holz}, \citenamefont {Wahab}, \citenamefont
  {Phoonthong}, \citenamefont {Cubero},\ and\ \citenamefont
  {Renzoni}}]{wickenbrock:2012}%
  \BibitemOpen
  \bibfield  {author} {\bibinfo {author} {\bibfnamefont {A.}~\bibnamefont
  {Wickenbrock}}, \bibinfo {author} {\bibfnamefont {P.~C.}\ \bibnamefont
  {Holz}}, \bibinfo {author} {\bibfnamefont {N.~A.~A.}\ \bibnamefont {Wahab}},
  \bibinfo {author} {\bibfnamefont {P.}~\bibnamefont {Phoonthong}}, \bibinfo
  {author} {\bibfnamefont {D.}~\bibnamefont {Cubero}}, \ and\ \bibinfo {author}
  {\bibfnamefont {F.}~\bibnamefont {Renzoni}},\ }\href@noop {} {\bibfield
  {journal} {\bibinfo  {journal} {Phys. Rev. Lett.}\ }\textbf {\bibinfo
  {volume} {108}},\ \bibinfo {pages} {020603} (\bibinfo {year}
  {2012})}\BibitemShut {NoStop}%
\bibitem [{\citenamefont {Anishchenko}\ \emph {et~al.}(2002)\citenamefont
  {Anishchenko}, \citenamefont {Neiman}, \citenamefont {Astakhov},
  \citenamefont {Vadiavasova},\ and\ \citenamefont
  {Schimansky-Geier}}]{anishchenko:2002}%
  \BibitemOpen
  \bibfield  {author} {\bibinfo {author} {\bibfnamefont {V.}~\bibnamefont
  {Anishchenko}}, \bibinfo {author} {\bibfnamefont {A.}~\bibnamefont {Neiman}},
  \bibinfo {author} {\bibfnamefont {A.}~\bibnamefont {Astakhov}}, \bibinfo
  {author} {\bibfnamefont {T.}~\bibnamefont {Vadiavasova}}, \ and\ \bibinfo
  {author} {\bibfnamefont {L.}~\bibnamefont {Schimansky-Geier}},\ }\href@noop
  {} {\emph {\bibinfo {title} {{Chaotic and Stochastic Processes in Dynamic
  Systems}}}}\ (\bibinfo  {publisher} {Springer},\ \bibinfo {address}
  {Berlin},\ \bibinfo {year} {2002})\BibitemShut {NoStop}%
\bibitem [{\citenamefont {Freund}\ \emph {et~al.}(2003)\citenamefont {Freund},
  \citenamefont {Schimansky-Geier},\ and\ \citenamefont
  {H{\"a}nggi}}]{freund:2003}%
  \BibitemOpen
  \bibfield  {author} {\bibinfo {author} {\bibfnamefont {J.~A.}\ \bibnamefont
  {Freund}}, \bibinfo {author} {\bibfnamefont {L.}~\bibnamefont
  {Schimansky-Geier}}, \ and\ \bibinfo {author} {\bibfnamefont
  {P.}~\bibnamefont {H{\"a}nggi}},\ }\href@noop {} {\bibfield  {journal}
  {\bibinfo  {journal} {Chaos}\ }\textbf {\bibinfo {volume} {13}},\ \bibinfo
  {pages} {225} (\bibinfo {year} {2003})}\BibitemShut {NoStop}%
\bibitem [{\citenamefont {Lindner}\ \emph {et~al.}(2004)\citenamefont
  {Lindner}, \citenamefont {Garcia-Ojalvo}, \citenamefont {Neiman},\ and\
  \citenamefont {Schimansky-Geier}}]{lindner:2004}%
  \BibitemOpen
  \bibfield  {author} {\bibinfo {author} {\bibfnamefont {B.}~\bibnamefont
  {Lindner}}, \bibinfo {author} {\bibfnamefont {J.}~\bibnamefont
  {Garcia-Ojalvo}}, \bibinfo {author} {\bibfnamefont {A.}~\bibnamefont
  {Neiman}}, \ and\ \bibinfo {author} {\bibfnamefont {L.}~\bibnamefont
  {Schimansky-Geier}},\ }\href@noop {} {\bibfield  {journal} {\bibinfo
  {journal} {Phys. Rep.}\ }\textbf {\bibinfo {volume} {392}},\ \bibinfo {pages}
  {321} (\bibinfo {year} {2004})}\BibitemShut {NoStop}%
\bibitem [{\citenamefont {Casado-Pascual}\ \emph
  {et~al.}(2005{\natexlab{b}})\citenamefont {Casado-Pascual}, \citenamefont
  {G{\'o}mez-Ord{\'o}{\~n}ez}, \citenamefont {Morillo}, \citenamefont
  {Lehmann}, \citenamefont {Goychuk},\ and\ \citenamefont
  {H{\"a}nggi}}]{casado1:2005}%
  \BibitemOpen
  \bibfield  {author} {\bibinfo {author} {\bibfnamefont {J.}~\bibnamefont
  {Casado-Pascual}}, \bibinfo {author} {\bibfnamefont {J.}~\bibnamefont
  {G{\'o}mez-Ord{\'o}{\~n}ez}}, \bibinfo {author} {\bibfnamefont
  {M.}~\bibnamefont {Morillo}}, \bibinfo {author} {\bibfnamefont
  {J.}~\bibnamefont {Lehmann}}, \bibinfo {author} {\bibfnamefont
  {I.}~\bibnamefont {Goychuk}}, \ and\ \bibinfo {author} {\bibfnamefont
  {P.}~\bibnamefont {H{\"a}nggi}},\ }\href@noop {} {\bibfield  {journal}
  {\bibinfo  {journal} {Phys. Rev. E}\ }\textbf {\bibinfo {volume} {71}},\
  \bibinfo {pages} {011101} (\bibinfo {year} {2005}{\natexlab{b}})}\BibitemShut
  {NoStop}%
\bibitem [{\citenamefont {Goychuk}\ \emph {et~al.}(2006)\citenamefont
  {Goychuk}, \citenamefont {Casado-Pascual}, \citenamefont {Morillo},
  \citenamefont {Lehmann},\ and\ \citenamefont {H{\"a}nggi}}]{goychuk:2006}%
  \BibitemOpen
  \bibfield  {author} {\bibinfo {author} {\bibfnamefont {I.}~\bibnamefont
  {Goychuk}}, \bibinfo {author} {\bibfnamefont {J.}~\bibnamefont
  {Casado-Pascual}}, \bibinfo {author} {\bibfnamefont {M.}~\bibnamefont
  {Morillo}}, \bibinfo {author} {\bibfnamefont {J.}~\bibnamefont {Lehmann}}, \
  and\ \bibinfo {author} {\bibfnamefont {P.}~\bibnamefont {H{\"a}nggi}},\
  }\href@noop {} {\bibfield  {journal} {\bibinfo  {journal} {Phys. Rev. Lett.}\
  }\textbf {\bibinfo {volume} {97}},\ \bibinfo {pages} {210601} (\bibinfo
  {year} {2006})}\BibitemShut {NoStop}%
\bibitem [{\citenamefont {Tessone}\ and\ \citenamefont
  {Toral}(2005)}]{tessone:2005}%
  \BibitemOpen
  \bibfield  {author} {\bibinfo {author} {\bibfnamefont {C.~J.}\ \bibnamefont
  {Tessone}}\ and\ \bibinfo {author} {\bibfnamefont {R.}~\bibnamefont
  {Toral}},\ }\href@noop {} {\bibfield  {journal} {\bibinfo  {journal} {Physica
  A}\ }\textbf {\bibinfo {volume} {351}},\ \bibinfo {pages} {106} (\bibinfo
  {year} {2005})}\BibitemShut {NoStop}%
\bibitem [{\citenamefont {Martins}\ \emph {et~al.}(2009)\citenamefont
  {Martins}, \citenamefont {Toral},\ and\ \citenamefont
  {Santos}}]{vaz_martins:2009}%
  \BibitemOpen
  \bibfield  {author} {\bibinfo {author} {\bibfnamefont {T.~V.}\ \bibnamefont
  {Martins}}, \bibinfo {author} {\bibfnamefont {R.}~\bibnamefont {Toral}}, \
  and\ \bibinfo {author} {\bibfnamefont {M.}~\bibnamefont {Santos}},\
  }\href@noop {} {\bibfield  {journal} {\bibinfo  {journal} {Eur. Phys. J. B}\
  }\textbf {\bibinfo {volume} {67}},\ \bibinfo {pages} {329} (\bibinfo {year}
  {2009})}\BibitemShut {NoStop}%
\bibitem [{\citenamefont {Tessone}\ and\ \citenamefont
  {Toral}(2009)}]{tessone:2009}%
  \BibitemOpen
  \bibfield  {author} {\bibinfo {author} {\bibfnamefont {C.~J.}\ \bibnamefont
  {Tessone}}\ and\ \bibinfo {author} {\bibfnamefont {R.}~\bibnamefont
  {Toral}},\ }\href@noop {} {\bibfield  {journal} {\bibinfo  {journal} {Eur.
  Phys. J. B}\ }\textbf {\bibinfo {volume} {71}},\ \bibinfo {pages} {549}
  (\bibinfo {year} {2009})}\BibitemShut {NoStop}%
\bibitem [{\citenamefont {Grossmann}\ \emph {et~al.}(1991)\citenamefont
  {Grossmann}, \citenamefont {Dittrich}, \citenamefont {Jung},\ and\
  \citenamefont {H{\"a}nggi}}]{grossmann:1991}%
  \BibitemOpen
  \bibfield  {author} {\bibinfo {author} {\bibfnamefont {F.}~\bibnamefont
  {Grossmann}}, \bibinfo {author} {\bibfnamefont {T.}~\bibnamefont {Dittrich}},
  \bibinfo {author} {\bibfnamefont {P.}~\bibnamefont {Jung}}, \ and\ \bibinfo
  {author} {\bibfnamefont {P.}~\bibnamefont {H{\"a}nggi}},\ }\href@noop {}
  {\bibfield  {journal} {\bibinfo  {journal} {Phys. Rev. Lett.}\ }\textbf
  {\bibinfo {volume} {67}},\ \bibinfo {pages} {516} (\bibinfo {year}
  {1991})}\BibitemShut {NoStop}%
\bibitem [{\citenamefont {Casati}\ \emph {et~al.}(1979)\citenamefont {Casati},
  \citenamefont {Chirikov}, \citenamefont {Ford},\ and\ \citenamefont
  {Izrailev}}]{casati:1979}%
  \BibitemOpen
  \bibfield  {author} {\bibinfo {author} {\bibfnamefont {G.}~\bibnamefont
  {Casati}}, \bibinfo {author} {\bibfnamefont {B.~V.}\ \bibnamefont
  {Chirikov}}, \bibinfo {author} {\bibfnamefont {J.}~\bibnamefont {Ford}}, \
  and\ \bibinfo {author} {\bibfnamefont {F.~M.}\ \bibnamefont {Izrailev}},\
  }\href@noop {} {\emph {\bibinfo {title} {{Stochastic Behaviour in Classical
  and Quantum Hamiltonian Systems}}}},\ edited by\ \bibinfo {editor}
  {\bibfnamefont {G.}~\bibnamefont {Casati}}\ and\ \bibinfo {editor}
  {\bibfnamefont {J.}~\bibnamefont {Ford}},\ \bibinfo {series} {Lecture Notes
  in Physics}, Vol.~\bibinfo {volume} {93}\ (\bibinfo  {publisher} {Springer},\
  \bibinfo {address} {Berlin},\ \bibinfo {year} {1979})\BibitemShut {NoStop}%
\bibitem [{\citenamefont {Moore}\ \emph {et~al.}(1994)\citenamefont {Moore},
  \citenamefont {Robinson}, \citenamefont {Bharucha}, \citenamefont
  {Williams},\ and\ \citenamefont {Raizen}}]{moore:1994}%
  \BibitemOpen
  \bibfield  {author} {\bibinfo {author} {\bibfnamefont {F.~L.}\ \bibnamefont
  {Moore}}, \bibinfo {author} {\bibfnamefont {J.~C.}\ \bibnamefont {Robinson}},
  \bibinfo {author} {\bibfnamefont {C.}~\bibnamefont {Bharucha}}, \bibinfo
  {author} {\bibfnamefont {P.~E.}\ \bibnamefont {Williams}}, \ and\ \bibinfo
  {author} {\bibfnamefont {M.~G.}\ \bibnamefont {Raizen}},\ }\href@noop {}
  {\bibfield  {journal} {\bibinfo  {journal} {Phys. Rev. Lett.}\ }\textbf
  {\bibinfo {volume} {73}},\ \bibinfo {pages} {2974} (\bibinfo {year}
  {1994})}\BibitemShut {NoStop}%
\bibitem [{\citenamefont {Robinson}\ \emph {et~al.}(1995)\citenamefont
  {Robinson}, \citenamefont {Bharucha}, \citenamefont {Moore}, \citenamefont
  {Jahnke}, \citenamefont {Georgakis}, \citenamefont {Niu}, \citenamefont
  {Raizen},\ and\ \citenamefont {Sundaram}}]{robinson:1995}%
  \BibitemOpen
  \bibfield  {author} {\bibinfo {author} {\bibfnamefont {J.~C.}\ \bibnamefont
  {Robinson}}, \bibinfo {author} {\bibfnamefont {C.}~\bibnamefont {Bharucha}},
  \bibinfo {author} {\bibfnamefont {F.~L.}\ \bibnamefont {Moore}}, \bibinfo
  {author} {\bibfnamefont {R.}~\bibnamefont {Jahnke}}, \bibinfo {author}
  {\bibfnamefont {G.~A.}\ \bibnamefont {Georgakis}}, \bibinfo {author}
  {\bibfnamefont {Q.}~\bibnamefont {Niu}}, \bibinfo {author} {\bibfnamefont
  {M.~G.}\ \bibnamefont {Raizen}}, \ and\ \bibinfo {author} {\bibfnamefont
  {B.}~\bibnamefont {Sundaram}},\ }\href@noop {} {\bibfield  {journal}
  {\bibinfo  {journal} {Phys. Rev. Lett.}\ }\textbf {\bibinfo {volume} {74}},\
  \bibinfo {pages} {3963} (\bibinfo {year} {1995})}\BibitemShut {NoStop}%
\bibitem [{\citenamefont {Ringot}\ \emph {et~al.}(2000)\citenamefont {Ringot},
  \citenamefont {Szriftgiser}, \citenamefont {Garreau},\ and\ \citenamefont
  {Delande}}]{ringot:2000}%
  \BibitemOpen
  \bibfield  {author} {\bibinfo {author} {\bibfnamefont {J.}~\bibnamefont
  {Ringot}}, \bibinfo {author} {\bibfnamefont {P.}~\bibnamefont {Szriftgiser}},
  \bibinfo {author} {\bibfnamefont {J.~C.}\ \bibnamefont {Garreau}}, \ and\
  \bibinfo {author} {\bibfnamefont {D.}~\bibnamefont {Delande}},\ }\href@noop
  {} {\bibfield  {journal} {\bibinfo  {journal} {Phys. Rev. Lett.}\ }\textbf
  {\bibinfo {volume} {85}},\ \bibinfo {pages} {2741} (\bibinfo {year}
  {2000})}\BibitemShut {NoStop}%
\bibitem [{\citenamefont {Astumian}\ and\ \citenamefont
  {H{\"a}nggi}(2002)}]{astumian:2002}%
  \BibitemOpen
  \bibfield  {author} {\bibinfo {author} {\bibfnamefont {R.~D.}\ \bibnamefont
  {Astumian}}\ and\ \bibinfo {author} {\bibfnamefont {P.}~\bibnamefont
  {H{\"a}nggi}},\ }\href@noop {} {\bibfield  {journal} {\bibinfo  {journal}
  {Phys. Today}\ }\textbf {\bibinfo {volume} {55}},\ \bibinfo {pages} {33}
  (\bibinfo {year} {2002})}\BibitemShut {NoStop}%
\bibitem [{\citenamefont {Reimann}(2002)}]{reimann:2002a}%
  \BibitemOpen
  \bibfield  {author} {\bibinfo {author} {\bibfnamefont {P.}~\bibnamefont
  {Reimann}},\ }\href@noop {} {\bibfield  {journal} {\bibinfo  {journal} {Phys.
  Rep.}\ }\textbf {\bibinfo {volume} {361}},\ \bibinfo {pages} {57} (\bibinfo
  {year} {2002})}\BibitemShut {NoStop}%
\bibitem [{\citenamefont {H{\"a}nggi}\ \emph {et~al.}(2005)\citenamefont
  {H{\"a}nggi}, \citenamefont {Marchesoni},\ and\ \citenamefont
  {Nori}}]{hanggi:2005}%
  \BibitemOpen
  \bibfield  {author} {\bibinfo {author} {\bibfnamefont {P.}~\bibnamefont
  {H{\"a}nggi}}, \bibinfo {author} {\bibfnamefont {F.}~\bibnamefont
  {Marchesoni}}, \ and\ \bibinfo {author} {\bibfnamefont {F.}~\bibnamefont
  {Nori}},\ }\href@noop {} {\bibfield  {journal} {\bibinfo  {journal} {Ann.
  Phys. (Leipzig)}\ }\textbf {\bibinfo {volume} {14}},\ \bibinfo {pages} {51}
  (\bibinfo {year} {2005})}\BibitemShut {NoStop}%
\bibitem [{\citenamefont {H{\"a}nggi}\ and\ \citenamefont
  {Marchesoni}(2009)}]{hanggi:2009}%
  \BibitemOpen
  \bibfield  {author} {\bibinfo {author} {\bibfnamefont {P.}~\bibnamefont
  {H{\"a}nggi}}\ and\ \bibinfo {author} {\bibfnamefont {F.}~\bibnamefont
  {Marchesoni}},\ }\href@noop {} {\bibfield  {journal} {\bibinfo  {journal}
  {Rev. Mod. Phys.}\ }\textbf {\bibinfo {volume} {81}},\ \bibinfo {pages} {387}
  (\bibinfo {year} {2009})}\BibitemShut {NoStop}%
\bibitem [{\citenamefont {Ponomarev}\ \emph {et~al.}(2009)\citenamefont
  {Ponomarev}, \citenamefont {Denisov},\ and\ \citenamefont
  {H{\"a}nggi}}]{ponomarev:2009}%
  \BibitemOpen
  \bibfield  {author} {\bibinfo {author} {\bibfnamefont {A.~V.}\ \bibnamefont
  {Ponomarev}}, \bibinfo {author} {\bibfnamefont {S.}~\bibnamefont {Denisov}},
  \ and\ \bibinfo {author} {\bibfnamefont {P.}~\bibnamefont {H{\"a}nggi}},\
  }\href@noop {} {\bibfield  {journal} {\bibinfo  {journal} {Phys. Rev. Lett.}\
  }\textbf {\bibinfo {volume} {102}},\ \bibinfo {pages} {230601} (\bibinfo
  {year} {2009})}\BibitemShut {NoStop}%
\bibitem [{\citenamefont {Casado-Pascual}\ \emph {et~al.}(2013)\citenamefont
  {Casado-Pascual}, \citenamefont {Cubero},\ and\ \citenamefont
  {Renzoni}}]{casado:2013}%
  \BibitemOpen
  \bibfield  {author} {\bibinfo {author} {\bibfnamefont {J.}~\bibnamefont
  {Casado-Pascual}}, \bibinfo {author} {\bibfnamefont {D.}~\bibnamefont
  {Cubero}}, \ and\ \bibinfo {author} {\bibfnamefont {F.}~\bibnamefont
  {Renzoni}},\ }\href@noop {} {\bibfield  {journal} {\bibinfo  {journal} {Phys.
  Rev. E}\ }\textbf {\bibinfo {volume} {88}},\ \bibinfo {pages} {062919}
  (\bibinfo {year} {2013})}\BibitemShut {NoStop}%
\bibitem [{\citenamefont {Cubero}\ \emph {et~al.}(2014)\citenamefont {Cubero},
  \citenamefont {Casado-Pascual},\ and\ \citenamefont {Renzoni}}]{cubero:2014}%
  \BibitemOpen
  \bibfield  {author} {\bibinfo {author} {\bibfnamefont {D.}~\bibnamefont
  {Cubero}}, \bibinfo {author} {\bibfnamefont {J.}~\bibnamefont
  {Casado-Pascual}}, \ and\ \bibinfo {author} {\bibfnamefont {F.}~\bibnamefont
  {Renzoni}},\ }\href@noop {} {\bibfield  {journal} {\bibinfo  {journal} {Phys.
  Rev. Lett.}\ }\textbf {\bibinfo {volume} {112}},\ \bibinfo {pages} {174102}
  (\bibinfo {year} {2014})}\BibitemShut {NoStop}%
\bibitem [{\citenamefont {Neumann}\ and\ \citenamefont
  {Pikovsky}(2002)}]{neumann:2002}%
  \BibitemOpen
  \bibfield  {author} {\bibinfo {author} {\bibfnamefont {E.}~\bibnamefont
  {Neumann}}\ and\ \bibinfo {author} {\bibfnamefont {A.}~\bibnamefont
  {Pikovsky}},\ }\href@noop {} {\bibfield  {journal} {\bibinfo  {journal} {Eur.
  Phys. J. B}\ }\textbf {\bibinfo {volume} {26}},\ \bibinfo {pages} {219}
  (\bibinfo {year} {2002})}\BibitemShut {NoStop}%
\bibitem [{\citenamefont {Cubero}\ and\ \citenamefont
  {Renzoni}(2012)}]{cubero:2012}%
  \BibitemOpen
  \bibfield  {author} {\bibinfo {author} {\bibfnamefont {D.}~\bibnamefont
  {Cubero}}\ and\ \bibinfo {author} {\bibfnamefont {F.}~\bibnamefont
  {Renzoni}},\ }\href@noop {} {\bibfield  {journal} {\bibinfo  {journal} {Phys.
  Rev. E}\ }\textbf {\bibinfo {volume} {86}},\ \bibinfo {pages} {056201}
  (\bibinfo {year} {2012})}\BibitemShut {NoStop}%
\bibitem [{\citenamefont {Morito}\ and\ \citenamefont
  {Salkin}(1980)}]{morito:1980}%
  \BibitemOpen
  \bibfield  {author} {\bibinfo {author} {\bibfnamefont {S.}~\bibnamefont
  {Morito}}\ and\ \bibinfo {author} {\bibfnamefont {H.~M.}\ \bibnamefont
  {Salkin}},\ }\href@noop {} {\bibfield  {journal} {\bibinfo  {journal} {Acta
  Inform.}\ }\textbf {\bibinfo {volume} {13}},\ \bibinfo {pages} {379}
  (\bibinfo {year} {1980})}\BibitemShut {NoStop}%
\bibitem [{\citenamefont {Cuesta}\ \emph {et~al.}(2013)\citenamefont {Cuesta},
  \citenamefont {Quintero},\ and\ \citenamefont
  {Alvarez-Nodarse}}]{cuesta:2013}%
  \BibitemOpen
  \bibfield  {author} {\bibinfo {author} {\bibfnamefont {J.~A.}\ \bibnamefont
  {Cuesta}}, \bibinfo {author} {\bibfnamefont {N.~R.}\ \bibnamefont
  {Quintero}}, \ and\ \bibinfo {author} {\bibfnamefont {R.}~\bibnamefont
  {Alvarez-Nodarse}},\ }\href {\doibase 10.1103/PhysRevX.3.041014} {\bibfield
  {journal} {\bibinfo  {journal} {Phys. Rev. X}\ }\textbf {\bibinfo {volume}
  {3}},\ \bibinfo {pages} {041014} (\bibinfo {year} {2013})}\BibitemShut
  {NoStop}%
\bibitem [{\citenamefont {Dalfovo}\ \emph {et~al.}(1999)\citenamefont
  {Dalfovo}, \citenamefont {Giorgini}, \citenamefont {Pitaevskii},\ and\
  \citenamefont {Stringari}}]{dalfovo:1999}%
  \BibitemOpen
  \bibfield  {author} {\bibinfo {author} {\bibfnamefont {F.}~\bibnamefont
  {Dalfovo}}, \bibinfo {author} {\bibfnamefont {S.}~\bibnamefont {Giorgini}},
  \bibinfo {author} {\bibfnamefont {L.~P.}\ \bibnamefont {Pitaevskii}}, \ and\
  \bibinfo {author} {\bibfnamefont {S.}~\bibnamefont {Stringari}},\ }\href@noop
  {} {\bibfield  {journal} {\bibinfo  {journal} {Rev. Mod. Phys.}\ }\textbf
  {\bibinfo {volume} {71}},\ \bibinfo {pages} {463} (\bibinfo {year}
  {1999})}\BibitemShut {NoStop}%
\bibitem [{\citenamefont {Pitaevskii}\ and\ \citenamefont
  {Stringari}(2003)}]{pitaevskii:2003}%
  \BibitemOpen
  \bibfield  {author} {\bibinfo {author} {\bibfnamefont {L.~P.}\ \bibnamefont
  {Pitaevskii}}\ and\ \bibinfo {author} {\bibfnamefont {S.}~\bibnamefont
  {Stringari}},\ }\href@noop {} {\emph {\bibinfo {title} {{Bose-Einstein
  Condensation}}}}\ (\bibinfo  {publisher} {Oxford University Press},\ \bibinfo
  {address} {New York},\ \bibinfo {year} {2003})\BibitemShut {NoStop}%
\bibitem [{\citenamefont {Poletti}\ \emph {et~al.}(2009)\citenamefont
  {Poletti}, \citenamefont {Benenti}, \citenamefont {Casati}, \citenamefont
  {H{\"a}nggi},\ and\ \citenamefont {Li}}]{poletti:2009a}%
  \BibitemOpen
  \bibfield  {author} {\bibinfo {author} {\bibfnamefont {D.}~\bibnamefont
  {Poletti}}, \bibinfo {author} {\bibfnamefont {G.}~\bibnamefont {Benenti}},
  \bibinfo {author} {\bibfnamefont {G.}~\bibnamefont {Casati}}, \bibinfo
  {author} {\bibfnamefont {P.}~\bibnamefont {H{\"a}nggi}}, \ and\ \bibinfo
  {author} {\bibfnamefont {B.}~\bibnamefont {Li}},\ }\href@noop {} {\bibfield
  {journal} {\bibinfo  {journal} {Phys. Rev. Lett.}\ }\textbf {\bibinfo
  {volume} {102}},\ \bibinfo {pages} {130604} (\bibinfo {year}
  {2009})}\BibitemShut {NoStop}%
\bibitem [{\citenamefont {Press}\ \emph {et~al.}(2007)\citenamefont {Press},
  \citenamefont {Teukolsky}, \citenamefont {Vetterling},\ and\ \citenamefont
  {Flannery}}]{num_rec}%
  \BibitemOpen
  \bibfield  {author} {\bibinfo {author} {\bibfnamefont {W.~H.}\ \bibnamefont
  {Press}}, \bibinfo {author} {\bibfnamefont {S.~A.}\ \bibnamefont
  {Teukolsky}}, \bibinfo {author} {\bibfnamefont {W.~T.}\ \bibnamefont
  {Vetterling}}, \ and\ \bibinfo {author} {\bibfnamefont {B.~P.}\ \bibnamefont
  {Flannery}},\ }\href@noop {} {\emph {\bibinfo {title} {{Numerical Recipes:
  The Art of Scientific Computing, 3rd ed.}}}}\ (\bibinfo  {publisher}
  {Cambridge University Press},\ \bibinfo {address} {New York},\ \bibinfo
  {year} {2007})\BibitemShut {NoStop}%
\end{thebibliography}%

\end{document}